\documentclass[useAMS,usenatbib]{mn2e}

\pdfoutput=1

\usepackage{graphicx,natbib,aas_macros,amssymb,lscape,epsfig,array,multirow,epsfig,rotating,lscape,enumerate}

\newcommand{\cN}[1]{\mathcal{N}}

\def\gsim{\;\rlap{\lower 2.5pt
 \hbox{$\sim$}}\raise 1.5pt\hbox{$>$}\;}
\def\lsim{\;\rlap{\lower 2.5pt
   \hbox{$\sim$}}\raise 1.5pt\hbox{$<$}\;}

\def\tablenotetext#1#2{
\@temptokena={\vspace{.5ex}{\noindent\llap{$^{#1}$}#2}\par}
\@temptokenb=\expandafter{\tblnote@list}
\xdef\tblnote@list{\the\@temptokenb\the\@temptokena}}

\title[On the Orbits of Low-mass Companions to White Dwarfs]{On the Orbits of Low-mass Companions to White Dwarfs and the Fates of the Known Exoplanets}

\author[Nordhaus \& Spiegel]{J.~Nordhaus$^{1,2,3}$\thanks{NSF Astronomy and Astrophysics Postdoctoral Fellow; E-mail: nordhaus@astro.rit.edu}
\&
D.~S.~Spiegel$^{4}$\\ 
$^1$ Center for Computational Relativity and Gravitation, Rochester Institute of Technology, Rochester, NY 14623, U.S.A.\\
$^2$ National Technical Institute for the Deaf, Rochester Institute of Technology, Rochester, NY 14623, U.S.A.\\
$^3$ Department of Physics and Astronomy, University of Rochester, Rochester, NY 14627, U. S. A.\\
$^4$ School of Natural Science, Institute for Advanced Study, Princeton, NJ 08540 U.S.A.}

\begin{document}
\pubyear{2013}
\maketitle
\label{firstpage}

\begin{abstract}
The ultimate fates of binary companions to stars (including whether the companion survives and the final orbit of the binary) are of interest in light of an increasing number of recently discovered, low-mass companions to white dwarfs (WDs).  In this {\it Letter}, we study the evolution of a two-body system wherein the orbit adjusts due to structural changes in the primary, dissipation of orbital energy via tides, and mass loss during the giant phases; previous studies have not incorporated changes in the primary's spin.  For companions ranging from Jupiter's mass to $\sim$0.3~$M_\odot$ and primaries ranging from 1--3~$M_\odot$, we determine the minimum initial semimajor axis required for the companion to avoid engulfment by the primary during post-main-sequence evolution, and highlight the implications for the ultimate survival of the known exoplanets.  We present regions in secondary mass and orbital period space where an engulfed companion might be expected to survive the common envelope phase (CEP), and compare with known M dwarf+WD short-period binaries.  Finally, we note that engulfed Earth-like planets cannot survive a CEP.  Detection of a first-generation terrestrial planet in the white dwarf habitable zone requires scattering from a several-AU orbit to a high-eccentricity orbit (with a periastron of $\sim$$R_\odot$) from which it is damped into a circular orbit via tidal friction, possibly rendering it an uninhabitable, charred ember.
\end{abstract}


\section{Introduction}
The ultimate fate of low-mass companions to main-sequence (MS) stars is of interest as substellar and stellar companions to intermediate-mass stars are plentiful \citep{Duquennoy:1991fk,Raghavan:2010qy,Wright:2011uq}.  Observationally, evidence for companions to evolved stars is varied.  In the early subgiant phase, a number of giant planets and giant planet candidates have been detected \citep{Johnson:2006rt,Johnson:2011ys}.  At the end of post-MS evolution, low-mass companions have been found in post-common-envelope, short-period orbits around subdwarfs and white dwarfs (see \citealt{Maxted:2006fj,Silvestri:2007mz,Charpinet:2011ly,Liu:2012ly,Rebassa-Mansergas:2012zr}; also Sec~\ref{Sec3}), and in long-period orbits around white dwarfs \citep{Farihi:2005fr,Farihi:2006vn,Farihi:2012mz}.

During post-MS evolution, dynamical interactions induced by radial expansion of the primary,  strong mass loss via stellar winds, and tidal interactions can occur.  Such processes have been previously studied for two-body systems \citep{Carlberg:2009uq,Villaver:2009qy,Nordhaus:2010zr,Mustill:2012lr} and many-body systems \citep{Veras:2011fj,Veras:2012kx,Kratter:2012yq,Perets:2012vn}.  In this {\it Letter}, we focus on two-body interactions.  We improve upon previous studies by including changes in the rotation rates of both bodies due to tidal dissipation and a varying moment of inertia.  During the subgiant and giant phases, the growth of the moment of inertia of the primary can cause companions within a few AU to switch from stable orbital configurations to being on the unstable side of the inner co-rotation point, thereby facilitating a plunge into their host stars \citep{Spiegel:2012lr}.  On the other hand, by tracking the spin-up of the primary, we find that higher-mass companions can retard their infall --- an effect not incorporated in previous studies.  For example, although \cite{Mustill:2012lr} include the evolution of the planet's spin (which is negligible at separations of a few AU\footnote{A jovian companion will experience a change in spin of $\lsim$1 part in $10^3 \times(a/{\rm 1~AU})^6$ during a Hubble time due to the stellar tide raised on it, where $a$ is the orbital separation.}), they neglect the changes in the primary's spin that drive the system's tidal evolution.  This is reasonable for the very-low-mass companions they consider, but calculating changes in the primary's spin is necessary for following the orbital evolution for higher-mass companions.

The structure of this paper is as follows.  In Section 2, we describe the physics and assumptions of our approach.  In Section 3, we present the minimum semimajor axis necessary to avoid engulfment as a function of companion mass and zero-age-main-sequence (ZAMS) primary mass.  For engulfed companions, we estimate whether or not the companion survives the common envelope and the corresponding orbital period at which it emerges.  In Section 4, we comment on the implications of finding a first-generation, terrestrial planet in the white dwarf habitable zone in the context of binary evolution.  We conclude in Section 5.

\section{Tidal Dissipation and Mass Loss}
\label{sec:2}
To study the evolution of our two-body system, we employ a tidal interaction model that couples the mass and radius of the primary star with the orbit of a secondary body.  Observational constraints on tides are difficult to achieve.  \cite{Zahn:1977jj,Zahn:1989lr} proposed a tidal theory based on turbulent viscosity, which we adopt in this work.  The theory was tested by introducing a dimensionless parameter $f$ that was then calibrated using eccentricity measurements of a sample of post-main-sequence binaries \citep{Verbunt:1995rt}.  Based on the divide between observed circularized and non-circularized systems, it has been argued that $f$ is constant and equal to unity\footnote{For a detailed description of the relation between $f$ and the tidal quality factor $Q'_\star$ \citep{Goldreich:1966qv} and the implications of the claim that $f=1$ see \S 2.3 and 4.1 of \cite{Nordhaus:2010zr}.} \citep{Verbunt:1995rt}.

The evolution of the semimajor axis, $a$, is as follows:
\begin{equation}
\frac{da}{dt} = \left(\frac{da}{dt}\right)_{\rm tides} + \left(\frac{da}{dt}\right)_{\rm mass-loss}.\label{eqn:dadt}
\end{equation}
where the first term represents the change in semimajor axis due to tidal dissipation (described below) and the second term represents the adiabatic change due to mass loss (i.e., $(da/dt)_{\rm mass-loss} \approx -a\dot{M_\star}/M_\star$).  Note that, while tidal dissipation occurs in both bodies, we assume that only the primary loses mass.  Since this material is lost from the system \citep{Spiegel:2012xt}, the second term is positive and acts to widen the orbit.  We take the orbit to be circular and refer the reader to \cite{Socrates:2012lr} for intricacies in modeling tidal dissipation in eccentric systems.

As the system moves toward synchronization, the change in rotation rate of each body is:
\begin{equation}
\frac{d\Omega_{\star,c}}{dt} = \left(\frac{d\Omega_{\star,c}}{dt}\right)_{\rm tides} + \left(\frac{d\Omega_{\star,c}}{dt}\right)_{\rm MoI}\label{eqn:domegadt},
\end{equation}
where $\Omega_\star$ and $\Omega_c$ are the spin rates of the primary and companion respectively.  The first term represents the change in the spin rate of a body due to tidal dissipation in its interior.  The second term affects the spin rate due to changes in the moment of inertia (MoI).    In this work, we assume that the companion's mass and moment of inertia do not change.  For a spin angular momentum loss rate of $\dot{S}_\star = (2/3) \dot{M}_\star R_\star^2 \Omega_\star$ from the primary, we have that:
\begin{equation}
\left(\frac{d\Omega_{\star}}{dt}\right)_{\rm MoI} = \Omega_{\star}\left\{\left(\frac{2}{3\alpha_{\star}}\right)\left(\frac{\dot{M}_{\star}}{M_{\star}}\right)- \frac{\dot{I}_{\star}}{I_{\star}}\right\},
\end{equation}
where  $I_\star$ is the primary's MoI, $\alpha_\star \equiv I_\star/(M_\star R_\star^2)$, and $\dot{M}_\star$ and $\dot{I}_\star$ are the time rates of change of the primary's mass and MoI.\footnote{If the wind is strongly magnetically coupled to the primary, Coriolis torques could render the situation somewhat more complicated than the model presented here.}  $I_\star$ is calculated from the stellar structure at each point of the evolution.  Our stellar models were computed using the {\it Modules for Experiments in Stellar Astrophysics (MESA)} code \citep{Paxton:2011lr} and produce $\sim$0.55--0.9-$M_\odot$~WDs for 1--3-$M_\odot$ ZAMS progenitors.

According to the viscous tide model of \citet{Zahn:1977jj,Zahn:1989lr}, the change in semimajor axis due to dissipation of orbital energy in the envelope of the primary is given as
\begin{eqnarray}
\left(\frac{da}{dt}\right)_{\rm tides} &=& -\frac{12ak_{2,\star}f}{\tau_{\star, {\rm conv}}}\left(\frac{M_{\rm \star,env}}{M_\star}\right)\left(\frac{M_c}{M_\star}\right)\left(1+\frac{M_c}{M_\star}\right)\nonumber\\
&&\times\left(\frac{R_\star}{a}\right)^8\left(1-\frac{\Omega_\star}{n}\right),
\end{eqnarray}
where $n$ is the orbital mean motion, $k_{2,\star}$ is the primary's tidal Love number which we assume to be unity.  The convective time of the primary is taken to be $\tau_{\star, {\rm conv}}\equiv \left(M_{\star, {\rm env}}R_\star^2/L_\star\right)^{1/3}$ such that $L_\star$ is the luminosity of the giant and $M_{\star, {\rm env}}$ is the mass of the convective envelope.  The change in rotation rate of the primary is then:
\begin{eqnarray}
 \left(\frac{d\Omega_\star}{dt}\right)_{\rm tides} = &&\frac{6nk_{2,\star}f}{\alpha_\star\tau_{\star, {\rm conv}}}\left(\frac{M_{\rm \star,env}}{M_\star}\right)\left(\frac{M_c}{M_\star}\right)^2\nonumber\\
 &&\times\left(\frac{R_\star}{a}\right)^6\left(1-\frac{\Omega_\star}{n}\right) - \Omega_\star\frac{\dot{I}_\star}{I_\star}.
 \label{eq:omega}
\end{eqnarray}
The equation for $\dot{\Omega}_c$ can be obtained by reversing the $\star$ and $c$ subscripts in Eq.~\ref{eq:omega}, although at $a$$\gsim1$ AU, $\dot{\Omega}_{\rm c}$ is negligible.

\section{Period Gaps}
\label{Sec3}
\begin{figure}
\begin{center}
\vspace{-0.1cm}
\includegraphics[width=6.9cm,angle=0,clip=true]{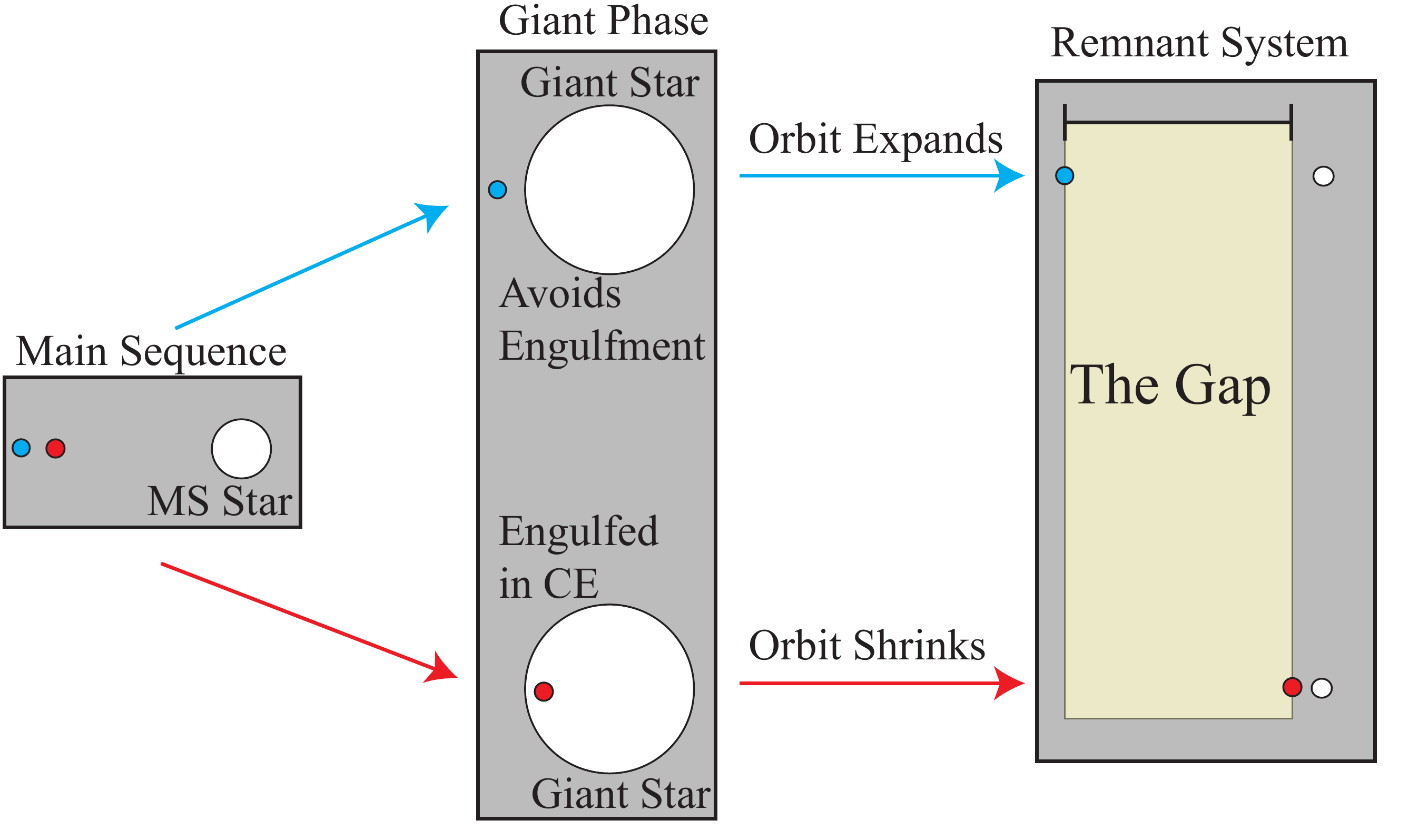}
\caption{The period gap.  A companion (blue circle) orbiting just exterior to the critical initial semimajor axis escapes engulfment such that the orbit expands.  A companion (red circle) orbiting just interior to the critical initial semimajor axis is engulfed such that the orbit shrinks during the CE phase.
\label{fig:cartoongap}}
\end{center}
\end{figure}

Due to the combined effects of tidally-induced orbital decay and mass-loss-induced orbital expanion, there should be a period gap in the distribution of low-mass companions to white dwarfs.  A cartoon schematic of this gap is presented in Fig.~\ref{fig:cartoongap}.  The outer edge of the gap for each binary configuration is given by the final orbital separation of the companion that escapes engulfment (blue circle in Fig.~\ref{fig:cartoongap}).  Since tidal torques drop off as a large negative power of orbital separation, most companions that escape engulfment experience essentially no tidal interactions.  Therefore, the outer boundary of the gap can be approximated as $a_{\rm i,crit} (M_{\rm \star,i} / M_{\rm wd})$ where $a_{\rm i,crit}$ is the minimum semi-major axis that escapes engulfment,\footnote{We define the companion to have been engulfed when it comes into Roche contact with the primary \citep{Kopal:1959fk,Eggleton:1983lr} and assume that its mass remains constant during the CEP \citep{Maxted:2006fj,Passy:2012fk}.}  $M_{\rm \star,i}$ is the primary's ZAMS mass and $M_{\rm wd}$ is the mass of the emergent white dwarf.  The inner edge is determined by the fate of the engulfed companion (red circle in Fig.~\ref{fig:cartoongap}) which will either emerge in a post-CE short-period orbit, or be destroyed during the CEP.

\begin{figure}
\begin{center}
\vspace{-0.2cm}
\includegraphics[width=7.5cm,angle=0,clip=true]{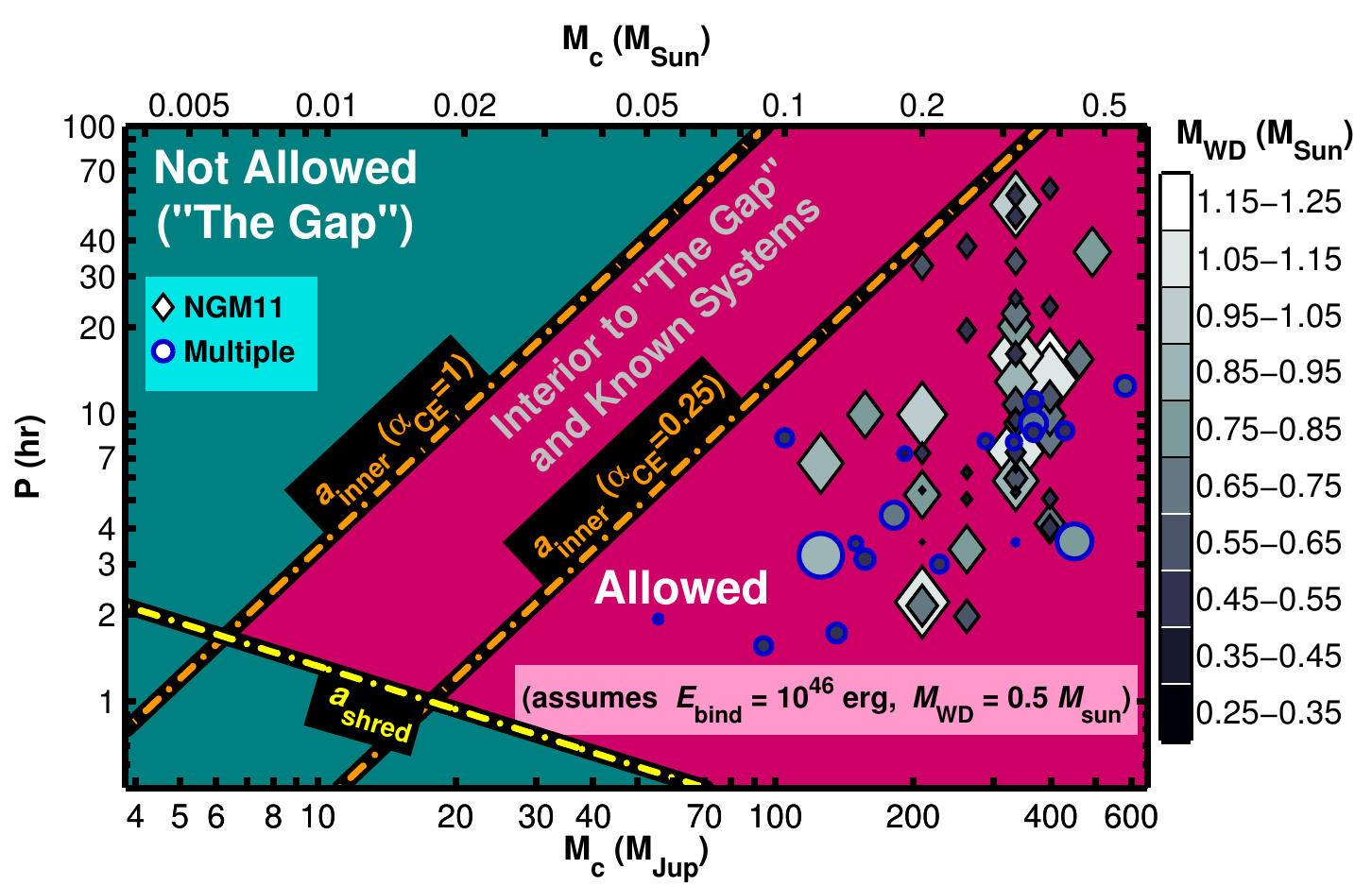}
\caption{The inner edge of the period gap.  Objects in the magenta region can survive CE evolution and emerge as short-period companions to WDs.  The $a_{\rm inner}$ boundaries are determined by assuming that $\alpha_{\rm CE}$ of the liberated orbital energy during inspiral is used to unbind the CE.  The $a_{\rm shred}$ boundary marks the location where companions tidally disrupt.  The known low-mass-short-period companions to WDs are marked with blue circles \citep{Parsons:2012lr,Parsons:2012fk,Parsons:2012qy,Haefner:2004kx,Parsons:2010yq,Pyrzas:2012uq,Law:2011fj,Tappert:2011lr,Badenes:2012yq,Pyrzas:2009fk,Nebot-Gomez-Moran:2009kx,van-den-Besselaar:2007vn,Maxted:2004rt,Maxted:2006fj,OBrien:2001ys} and black diamonds \citep{Nebot-Gomez-Moran:2011xx}.  Greyscale and marker size indicate companion mass.
\label{fig:Inner_edge}}
\end{center}
\end{figure}

The inner edge of the period gap is presented in Fig.~\ref{fig:Inner_edge}.  The ``allowed region" is marked in magenta.  The $a_{\rm inner}$ boundaries (orange dash-dot lines) are determined by calculating the semimajor axis where a fraction $\alpha_{\rm CE}$ \citep{Livio:1988qy,Soker:2013kx} of the liberated orbital energy during inspiral is sufficient to eject the envelope assuming that the envelope binding energy, $E_{\rm bind}=10^{46}$ erg and the mass of the WD is $M_{\rm WD} = 0.5$ $M_\odot$.\footnote{In a small fraction of systems ($\lesssim$1\%?), a planetary companion might be scattered to a high-eccentricity orbit that damps into a short-period orbit in ``the gap" region of Fig.~\ref{fig:cartoongap}, similar to the formation mechanism of hot Jupiters hypothesized by \citet{Fabrycky:2007ve}.}  An additional constraint for surviving the CE phase is that the companion does not tidally disrupt during inspiral \citep{Nordhaus:2006oq,Nordhaus:2011lq,Spiegel:2012lr}.  This $a_{\rm shred}$ boundary (yellow dash-dot line) is shown in Fig.~\ref{fig:Inner_edge}.  Additionally, the properties of the known short-period, low-mass companions to white dwarfs are marked by circles and diamonds in Fig.~\ref{fig:Inner_edge}.  Attempts have been made to calculate $\alpha_{\rm CE}$ from 3-D simulations \citep{Ricker:2012yq,Passy:2012vn}, and from observations \citep{Zorotovic:2010fr,De-Marco:2011uq} but the large range of scales involved in a CE-inspiral makes this a difficult calculation and there is, as yet, no consensus.  In Fig.~\ref{fig:Inner_edge}, we present $a_{\rm inner}$ for $\alpha_{\rm CE}$ values of 0.25  \citep{Zorotovic:2010fr} and 1.

\begin{figure*}
\begin{center}
\includegraphics[height=7.32cm, angle=0,clip=true]{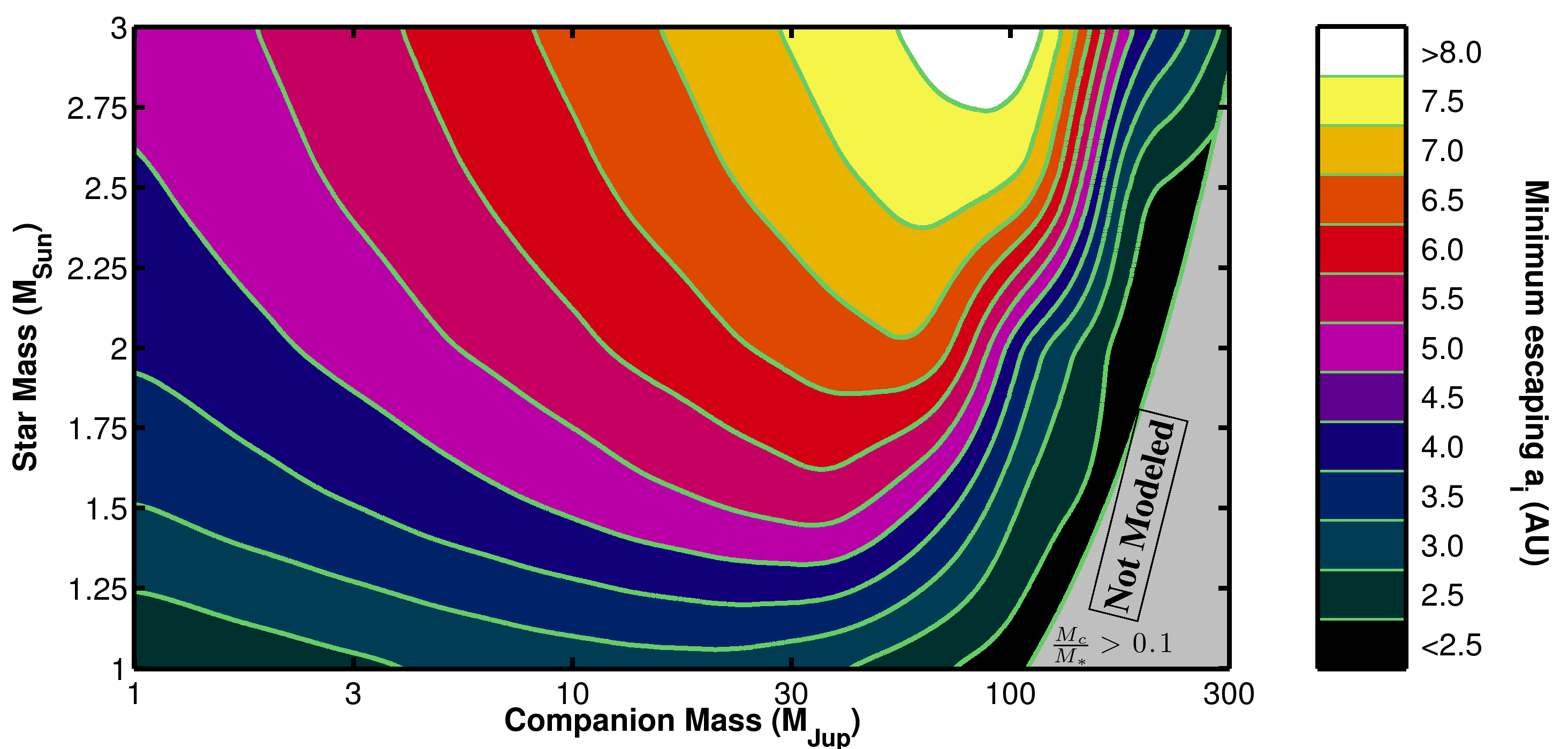} 
\caption{The minimum separation required to escape engulfment ($a_{\rm i,crit}$) as a function of companion mass and ZAMS primary mass under the assumption that $f=1$ (Sec.~\ref{sec:2}).  The outer bound of the gap can be estimated as $a_{\rm i,crit} \times(M_\star / M_{\rm WD})$, where $M_\star$ is the primary's ZAMS mass (the ordinate in this figure) and $M_{\rm WD}$ is the remnant white dwarf mass.  This estimate suggests that the outer boundary of the gap ranges from $\sim$$4$--$40$~AU for the primaries and companions considered in this figure.  Regions where the companion's mass exceeds one tenth of the primary's ZAMS mass are shown in grey and excluded from the calculations.
\label{fig:aouter}}
\end{center}
\end{figure*}

Figure~\ref{fig:aouter} shows the minimum initial semimajor axis required to escape engulfment as a function of primary and companion masses.  Employing the \citeauthor{Verbunt:1995rt} ($f=1$) calibration of the \citet{Zahn:1977jj,Zahn:1989lr} tidal formalism leads to very strong tides during the post-MS, with $Q'_\star$ values as low as $\sim$$10^{1}$--$10^{3}$ at the time of plunge \citep{Nordhaus:2010zr}.  The values of the contours in Fig.~\ref{fig:aouter} are sensitive to the maximum radius of the stellar model; changes in the stellar model that influence the maximum radius correspondingly influence $a_{\rm i, crit}$.  The discovery of companions near the outer edge of the gap would constrain the nature of tidal dissipation and the late stages of stellar evolution.  

Note that in Fig.~\ref{fig:aouter}, there is a decrease in the semimajor axis required to escape engulfment for companions greater than $\sim$$60$ $M_J$.  This is because at equal separation from the star, a more massive companion torques the star more strongly in proportion to the square of its mass.  As the star ascends the asymptotic giant branch (AGB), a sufficiently massive companion can halt its infall by synchronizing the primary.  Therefore, for a very massive companion to be engulfed, its initial separation needs to be smaller than that of a less massive companion.  Extremely massive companions can tidally transfer enough angular momentum to significantly spin up the primary.  In such cases, enhanced mass-loss and deformation of the primary may occur.  Since these processes are not modeled in our calculations, we exclude this region of parameter space (shown in grey in Fig.~\ref{fig:aouter}).

\begin{figure}
\begin{center}
\includegraphics[height=6.8cm, angle=0,clip=true]{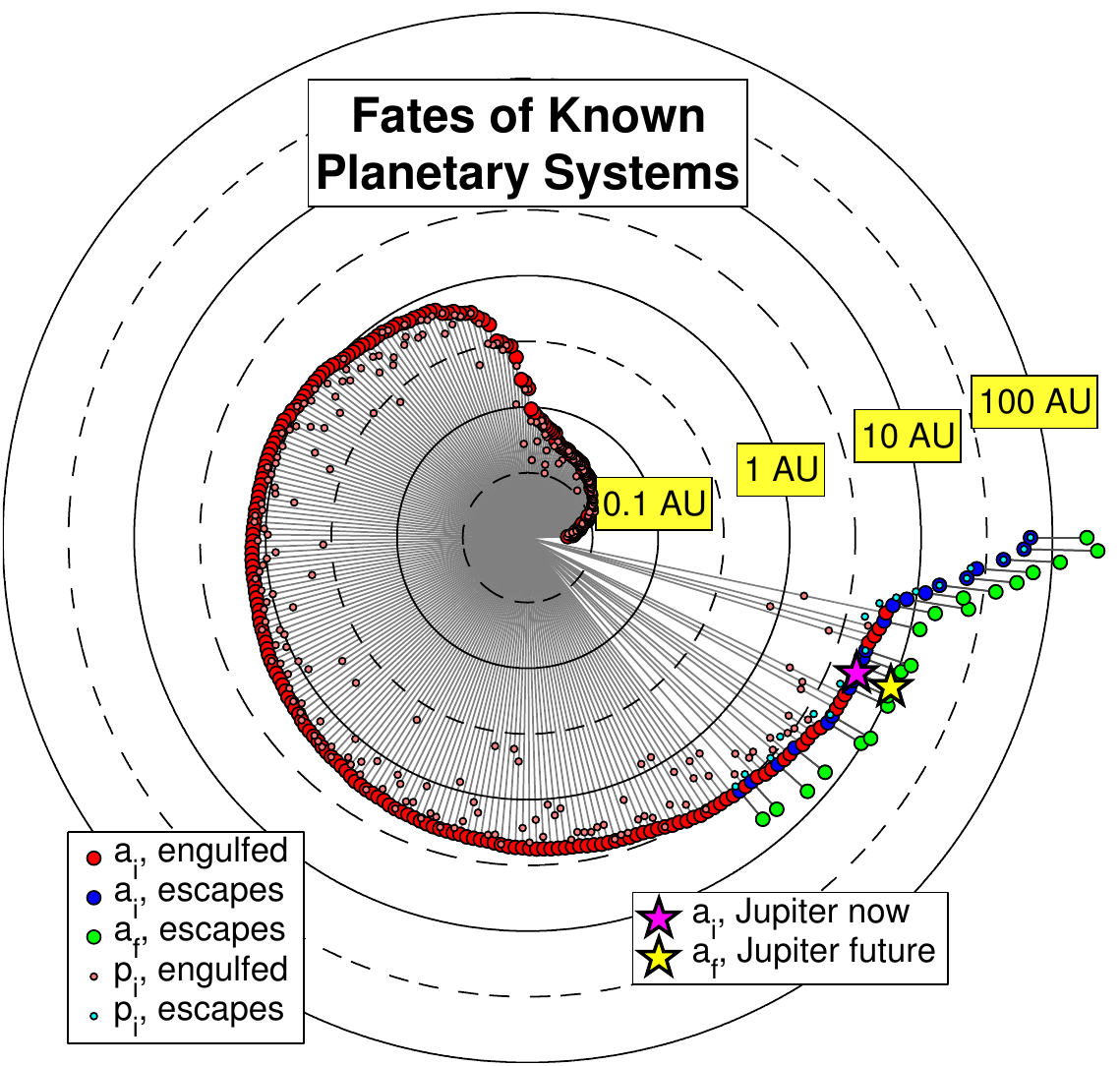} 
\caption{The fates of known planetary systems.  Red and blue dots indicate current semimajor axes for each of the $\sim$300 known exoplanets with masses at least Jupiter's and host-star masses greater than or equal to the Sun's.  A red dot indicates that the planet will be engulfed (according to Fig.~\ref{fig:aouter}) while a blue dot indicates that the planet will escape enfulgment; green dots indicate final circum-WD separations.  Small light-red (light-blue) dots indicate current periastra of planets that will be (avoid being) engulfed.  Current and final orbital semimajor axes of Jupiter are shown with magneta and yellow stars, respectively.  Among the currently known planetary systems, none will evolve to have circum-WD jovian planets within $\sim$6.4 AU.  
\label{fig:futures}}
\end{center}  
\end{figure}

Figure \ref{fig:futures} shows the fates of the known planets with masses greater than or equal to Jupiter's and host-star masses greater than or equal to the Sun's (data from the Open Exoplanet Catalogue: \citealt{Rein:2012fk}).  For each system, we calculate the joint stellar and tidal evolution to determine whether the planet will be engulfed in the post-MS stages.  For those objects that avoid engulfment, we use the WD initial-final mass relation of \citet{Catalan:2008lr} to determine the post-MS orbit expansion.  No known systems will evolve to have circum-WD jovian planets within $\sim$6.4 AU.  Relatively few known exoplanets will survive to orbit a WD.  Predictions of the demographics of WD planets will improve with a better accounting of the distribution of MS planets in the 3-10 AU range, which should be possible with micro-lensing studies with WFIRST \citep{Green:2012xx}.

\section{Are There Habitable Earths Around White Dwarfs?}
\label{sec:WDearths}

The annulus around a white dwarf that is amenable to habitable
climates \citep{Kasting:1993vn,Spiegel:2008rt} might be at an orbital separation of $\sim$1~$R_\odot$
\citep{Agol:2011ys,Fossati:2012fr,Loeb:2013xx}. How could a terrestrial
planet end up in such a potentially habitable orbit?

There are two ways that a first-generation planet could end up in a close
orbit around a white dwarf.  It either (a) survives a common envelope stage with a sub-giant, red-giant, or AGB
star, in which the companion is engulfed and inspirals inward until its
plunge is arrested before tidal disruption very close to the
degenerate core (at distances of $\lsim$0.3$R_\odot$), or (b) migrates
inward after the white dwarf has finished forming and expelled its
envelope.

Scenario (a) --- CE survival --- is implausible for a companion that
is less than $\sim$6~$M_J$ (see Fig.~\ref{fig:Inner_edge}),
because such low-mass companions lack enough energy to unbind even an evolved star's envelope, a necssary but not sufficient condition for CE survival \citep{Nordhaus:2010zr,Spiegel:2012lr}.  Furthermore, the inspiral
accelerates as the companion moves inward \citep{Nordhaus:2006oq,Nordhaus:2007il},
which means that for the star to unbind its own envelope right
when the companion is at 1~$R_\odot$ would require extreme fine tuning.
Even if a low-mass companion were to arrest its inspiral at $\sim$1~$R_\odot$, the high temperatures that it would encounter deep in the envelope of the
primary ($\sim$10$^6$~K) might pose a severe threat to its subsequent habitability or to its very survival \citep{Villaver:2007lr}.


Scenario (b) --- post-WD orbital evolution --- might occur, but could render planets that experience such a
process uninhabitable by life as we know it.  Indeed,
\citet{Zuckerman:2010mz} and others have found evidence of tidally
shredded asteroids accreting metals onto WDs, indicating that processes
can occur in post-main-sequence planetary systems that excite extreme
eccentricities among low-mass particles.  Via the Kozai mechanism
\citep{Kozai:1962ly,Fabrycky:2007ve,Katz:2011ul,Socrates:2012qf,Naoz:2012pd,Shappee:2012bh}, a massive
outer body can drive an inner companion to arbitrarily high
eccentricities, such that asteroids tidally disrupt very
near the WD.  Similar processes certainly might drive an Earth-mass
planet from a several-AU orbit (far enough out that it avoided being
engulfed during the AGB phase) to a high-eccentricity orbit with a
periastron of 0.5~$R_\odot$, from which tidal friction would damp it to
a 1-$R_\odot$ circular orbit.  However, if this happens, a large amount
of orbital energy must be dissipated as heat:
\begin{equation}
\Delta E_{\rm orb}  \sim -3 \times 10^{42} {\rm~ergs} \times \left(\frac{M_{\rm WD}}{0.5 M_\odot} \right)\left( \frac{M_p}{M_\oplus} \right) \left( \frac{a_{\rm final}}{R_\odot} \right)^{-1} \,
\end{equation}
where $M_p$ is the planet's mass.  Dissipation of this heat over a circularization timescale of $t_{\rm circ}\sim10^6$ years
corresponds to a tidal heat flux of $\sim$$\left(2\times
10^7 {\rm~W~m^{-2}}\right)\times\left(t_{\rm circ}/{\rm 1 Myr}\right)^{-1}$, or an average cooling temperature of $\sim$$\left(4000~K\right)\times\left(t_{\rm circ}/{\rm 1 Myr}\right)^{-1/4}$, which could be catastrophic for the habitability of such a planet (see \citealt{Barnes:2012fk} and \citealt{Barnes:2012bh}, who considered the adverse effects on habitability of much lower tidal power).
Such a migrated terrestrial planet might exist as a charred ember in the habitable
zone.  Though late-time delivery of volatiles might reintroduce water to a desiccated WDHZ planet, the near-daily cometary impacts on the Sun \citep{Marsden:2005xx} imply that an Earth at a 1-$R_\odot$ orbital radius could be subject to cometary hits at $\sim$$(R_\oplus/R_\odot)^2$ times this rate or once per several decades.  These impacts would have $\sim$100 times the specific energy of what wiped out the dinosaurs.  If large, these impactors could sterilize the worlds; if small, they would deliver very little water.

\section{Conclusions}
We have studied the effect of post-MS evolution on the orbits of substellar and M-dwarf companions.  For each combination of primary and companion mass, we evolve the system from the ZAMS through the end of the post-MS.  The evolution is governed by tidal dissipation, mass loss from the system and structural changes in the primary.  For each binary configuration, we determine the initial semimajor axis required to escape engulfment.  We note that terrestrial planets cannot survive engulfment.  Therefore, a first generation terrestrial planet in the WDHZ must have experienced an enormous tidal flux and, though formally habitable, might be inhospitable.  Interestingly, no currently known jovian planets will evolve to have circum-WD orbits within $\sim$6.4~AU.

\vspace{0.3cm}
\noindent{\uppercase{\bf Acknowledgments}}\\
JN is supported by NSF AAP Fellowship AST-1102738 and by NASA HST grant AR-12146.04-A.  DSS gratefully acknowledges support from NSF grant AST-0807444 and the Keck Fellowship, and the Friends of the Institute.  We thank P. Arras, A. Nebot G{\'o}mez-Mor{\'a}n, D. Maoz, A. Loeb, B. G\"{a}nsicke, N. Madappatt, O. De Marco, E. Agol, A. Socrates, R. Fernandez, JC Passy, S. Tremaine and J. Goodman for stimulating discussions.

\vspace{-0.4cm}
\bibliography{Gap}
\bibliographystyle{astron}

\end{document}